\documentstyle[12pt]{article}
\newcommand{\simlt}{\,\,\raise2pt\hbox{$<$}\llap{\lower3pt\hbox{$\sim$}}\,\,}
\newcommand{\simgt}{\,\,\raise2pt\hbox{$>$}\llap{\lower3pt\hbox{$\sim$}}\,\,}
\long\def\comment#1{ }
\def\alarm#1#2{#1_{\lower2pt\hbox{$\scriptstyle {\rm #2}$}}}
\def\ala#1#2{#1_{\lower2pt\hbox{$\scriptstyle {#2}$}}}
\def\alasrm#1#2{#1_{\lower1.5pt\hbox{$\scriptscriptstyle {\rm #2}$}}}
\def\alas#1#2{#1_{\lower1.5pt\hbox{$\scriptscriptstyle {#2}$}}}
\def\pt{\ala pT}
\def\pT{\ala pT}
\def\as{\alpha_{\rm s}}
\def\MRSDZ{MRS D$0^\prime$}
\def\DZ{D$0^\prime$}
\def\MRSDM{MRS D$-^\prime$}
\def\DM{D$-^\prime$}
\def\muR{\ala{\mu}{R}}
\def\muF{\ala{\mu}{F}}
\def\xF{\ala{x}{F}}
\newcommand{\yourtitle}[1]{
\mbox{}\\
\vskip 4\baselineskip
{\bf\noindent #1}\\ }
\newcommand{\youraddress}[1]{
\noindent\mbox{}\hspace{1in}\parbox[t]{4.0in}{#1}\\ }
\newcommand{\yournames}[1]{
\mbox{}\\
\mbox{}\\
\vskip-0.8cm
\noindent\mbox{}\hspace{1in}{#1}\\ }
\newcommand{\yourabstract}[1]{
\mbox{}\\
\mbox{}\\
{\bf\noindent Abstract}\\
\begin{center}
\mbox{}\parbox[t]{5.in}{#1}
\end{center} }
\newcommand{\yoursection}[1]{
\vskip 2\baselineskip
{\bf\noindent #1}\\
\mbox{}\\
\vspace{-0.19in}}

\begin{document}
\vskip-2cm
\rightline{HU-TFT-95-14}
\vskip1.2cm
\null\vskip-1cm

\yourtitle{PRODUCTION OF DRELL--YAN PAIRS IN HIGH ENERGY
NUCLEON--NUCLEON COLLISIONS }
\yournames{S.~Gavin and R.~Kauffman}
\youraddress{Department of Physics, \\
Brookhaven National Laboratory,\\
Upton, New York 11973, USA}
\yournames{S.~Gupta}
\youraddress{Theory Group, Tata Institute of Fundamental Research,\\
Homi Bhabha Road, Bombay 400 005, India}
\yournames{P.~V.~Ruuskanen}
\youraddress{Institute for Theoretical Physics, \\
P.O.Box 9, FIN-00014, University of Helsinki, Finland}
\yournames{D.~K.~Srivastava}
\youraddress{Variable Energy Cyclotron Centre, \\
1/AF Bidhan Nagar,  Calcutta 700 064, India}
\yournames{R.~L.~Thews}
\youraddress{Department of Physics, University of Arizona,\\
Tucson, Arizona 85721, USA}
\yourabstract{
We compute cross sections for the Drell-Yan process in N--N collisions at
next-to-leading order in $\alpha_s$.  The mass, rapidity, transverse momentum,
and angular dependence of these cross sections are presented.  An estimate of
higher order corrections is obtained from next-to-next-to-leading order
calculation of the mass distribution.  We compare the results with some of the
existing data to show the quality of the agreement between calculations and
data.  We present predictions for energies which will become available at the
RHIC and LHC colliders.  Uncertainties in these predictions due to choices
of scale, scheme and parton distribution are discussed.
}
\clearpage
\yoursection{INTRODUCTION}

The aim of this study is to provide a systematic survey of theoretical
predictions for the Drell--Yan process \cite{DY,Neerven} in
nucleon--nucleon collisions at energies relevant to ion--ion
experiments at RHIC and LHC, and to discuss confidence limits for
these predictions. In an accompanying article, Van Neerven has reviewed
the theory of the Drell--Yan process, emphasizing the dependence of
the production rate on the dilepton's mass $M$ and rapidity $y$.  We
present calculations of the $M$ and $y$ distributions using standard
perturbative QCD.  To supplement these calculations, we provide a
skeletal theoretical discussion to fix the notation and identify the
uncertainties.  In addition, we study the experimentally-relevant
transverse momentum and angular distributions of the dileptons.  These
topics are treated in separate subsections, since one must go beyond
perturbation theory to compute these distributions.

Our predictions for ${\rm d}\sigma/{\rm d}M{\rm d}y$ are based on a
perturbative analysis of the underlying partonic processes to order
$\alpha_{\rm s}$ \cite{DY1,Neerven1,Neerven2,Neerven3}. Results for
${\rm d}\sigma/{\rm d}M$ are reported to order $\alpha_{\rm s}^2$.  We
find that the perturbative corrections grow as $M$ decreases.  From
the point of view of the heavy ion physics, the mass region from 3 to
10 GeV is of most interest.  The relative magnitude of the
$O(\alpha_{\rm s}^2)$ correction in this range sets one limit on our
confidence in the applicability of perturbation theory.

At fixed order the calculated cross sections depend on the
renormalization scale $\muR$, the factorization scale $\muF$,
and the regularization scheme.  The form of the renormalized
hard-scattering matrix elements and the definition of the parton
distributions are specified by the regularization scheme; DIS and
$\overline{\rm MS}$ schemes are widely used.  The physical quantities
such as $\alpha_{\rm s}$ that enter the matrix elements are
defined at the scale $\muR$, while the parton distributions are
set at $\muF$.  Although these scales are related to the momentum
transfer $Q$, the precise relation is process dependent and not
unique.  The standard parton distribution sets have been obtained
assuming $\muF =\muR\equiv \mu$ \cite{PDFLIB,MRS,GRV}.

The scale and scheme dependence of our calculations provides an
additional measure of the accuracy of the perturbative description at
the given order.  From the standpoint of perturbation theory, the
choices of scales and scheme are arbitrary --- varying these choices
introduces corrections at the next order in $\alpha_{\rm s}$.
However, changing the scales and scheme in practice alters the
numerical predictions for collisions in the kinematic range
relevant to heavy ion experiments.  In this work we discuss results
for the DIS and $\overline{\rm MS}$ schemes and vary $\mu$ to test
the scale dependence.

Confidence in our predictions at the LHC heavy ion energy $\sqrt{s}\sim
5.5$~$A\cdot$TeV is further limited by current experimental uncertainties in
the parton distributions.  Specifically, the production of dileptons with $M<
10$~GeV in nucleon--nucleon collisions at this energy probes the parton
distributions at Bjorken $x < 10^{-2}$.  This region is accessible only
to the ongoing experiments at HERA \cite{HERA}. Consequently the differences
between the various parton distribution sets is largest in this region
\cite{PDFLIB,MRS,GRV}.  We base our predictions on computations using
state--of--the--art parton distribution sets that are consistent with the
current (1994) HERA data.  To illustrate the maximum uncertainty in these
predictions, we compare these results to calculations using a recent set that
does not exhibit the `low-$x$ rise' seen by HERA \cite{HERA}, \MRSDZ.
As the experiments accumulate data, these uncertainties will be reduced,
thereby enabling more refined predictions before the start of the LHC program.

We outline the theory used to study the
mass, rapidity, transverse momentum and angular distribution of the
dileptons in the next section.  In the following section we
compare our results to data and obtain predictions for RHIC and
LHC.  Results for ${\rm d}\sigma/{\rm d}M{\rm d}y$ are obtained using
a code provided by W.~van Neerven and P.~Rijken.   Transverse momentum
spectra and angular distributions are obtained following
Refs.~\cite{russ} and \cite{bob} respectively.  The computation of
these distributions --- and the $\pT$ spectrum in
particular --- requires a partial resummation of the perturbation
series together with nonperturbative input not contained in standard
parton distributions.  The methods and uncertainties specific to these
processes are discussed in the appropriate subsections.

\yoursection{THEORETICAL BACKGROUND}

We now discuss the calculation of the Drell--Yan cross section in
perturbative QCD.  Our goal here is to outline the theory so that the
reader can make use of our numerical results without extensive
recourse to the literature. We provide a list of essential references,
but those who are interested in a more detailed discussion of should
consult the accompanying article of van Neerven \cite{Neerven}.

\yoursection{Mass distributions}

The lowest order contribution to the Drell Yan process is
quark--antiquark annihilation into a lepton pair.  The annihilation
cross section can be obtained from the $e^+e^-\to\mu^+\mu^-$ cross
section by including the color factor 1/3 and the charge factor
$e_q^2$ for the quarks. Since the variation of the center--of--mass
energy $\sqrt{\hat s}$ of the incoming quark and antiquark leads to
pairs of different masses, it is useful to consider a cross section
that is differential in the mass $M$ of the pair:
\begin{equation}
\frac{{\rm d}\hat\sigma} { {\rm d}M^2}
= e_q^2\hat\sigma_0\delta(\hat s-M^2),
\qquad \hat\sigma_0=\frac{4\pi\alpha^2}{9M^2}
\end{equation}
The four--momenta of the incoming partons are expressed in terms of the
momentum fractions of the colliding hadrons as
\begin{equation}
p_1=\frac{\sqrt s}{2}(x_1,0,0,x_1) \qquad\qquad
p_2=\frac{\sqrt s}{2}(x_2,0,0,-x_2),
\end{equation}
where $\sqrt s$ is the center--of--mass energy of the hadrons.
It follows that ${\hat s}=x_1x_2s$.

The lowest order hadronic cross section is now obtained by folding in the
initial state quark and antiquark luminosities determined by the parton
distribution functions:
\begin{equation}
\frac{{\rm d}\sigma} { {\rm d}M^2} = \hat\sigma_0\int_0^1dx_1dx_2
\delta(x_1x_2s-M^2)\sum_k e_k^2[q_k(x_1,\mu)\bar q_k(x_2,\mu)
+(1\leftrightarrow 2)].
 \end{equation}
More precisely, the distributions $q$ and $\bar q$ give the number densities
of quarks and antiquarks at momentum fraction $x$ and factorization scale
$\mu$ which is of the order of $M$, the only scale entering the
calculation of the mass distribution.

The momentum fractions of the incoming partons which contribute to the LO
cross section can be expressed in terms of
the rapidity of the pair, $y$, and a scaling variable $\tau=M^2\slash s$ as
\begin{equation}
x_{01}=\sqrt\tau e^y, \qquad x_{02}=\sqrt\tau e^{-y}.
\end{equation}
Using $y=(1/2)\ln(x_{01}/x_{02})$, we write the double-differential cross
section %
\begin{equation}
\left( M^2\frac{{\rm d}\sigma} {{\rm d}y{\rm d}M^2}\right)_{\rm Born} =
{\hat\sigma}_0\tau
\sum_k e_k^2[q_k(x_{01},\mu)\bar q_k(x_{02},\mu)+(1\leftrightarrow 2)] =
F(\tau,\mu), \end{equation}
exhibiting a scaling behavior in $\tau$ at leading order (apart from the
logarithmic dependence on the factorization scale $\mu$).

The inclusive lepton pair cross section also includes contributions
from processes in which the final state contains partons in
addition to the lepton pair.  These processes are higher order in the
QCD coupling $\alpha_{\rm s}$.  Perturbative QCD provides a
systematic way to calculate order by order in $\alpha_{\rm s}$,
the contributions from such processes as well as from those with
virtual quanta. Graphs for the next to leading order processes
include Compton, annihilation, and vertex corrections.
The complete next to leading order cross section is \cite{DY1}
\begin{eqnarray}
\left(\frac{{\rm d}\sigma}{{\rm d}y{\rm d}M^2}\right)_{\rm NLO}
&\!\!\!\!\!\!
=&\!\!\frac{\hat{\sigma}_0}{s}
\int_0^1\,dx_1\,dx_2\,dz\,
\delta(x_1x_2z-\tau)\delta(y- \frac{1}{2}\ln\frac{x_1}{x_2}) \nonumber\\
&\times&\left\{\left[\sum_{k}e_k^2(q_k(x_1)\bar{q}_k(x_2)+
[1\leftrightarrow 2])\right]\left[\delta(1-z)+\frac{\alpha_{\rm s}(\mu)}
 {2\pi}f_q(z)\right]\right. \label{eq:nlo}  \\
&+&\left.\left[\sum_{k}e_k^2(g(x_1)(q_k(x_2)+\bar{q}_k(x_2))+
[1\leftrightarrow 2])\right]\left[\frac{\alpha_{\rm s}(\mu)}
{2\pi}f_g(z)\right]\right\},\nonumber
\end{eqnarray}
where the $g$ and $q_k$ are evaluated at the scale $\mu$.
The correction terms in the DIS regularization scheme 
are
\begin{eqnarray}
f_q(z)&=&C_F\left[ \delta(1-z)\left(1+\frac{4\pi^2}{3}\right)-6-4z+
\left(\frac{3}{1-z}\right)_+
+ 2(1+z^2)\left(\frac{\ln(1-z)}{1-z}\right)_+\right]~,\nonumber\\
f_g(z)&=&\frac{1}{2}\left[(z^2+(1-z)^2\ln(1-z)+\frac{3}{2}-5z+\frac{9}{2}z^2
\right].
\end{eqnarray}
Similar terms can be written down for the $\overline{MS}$ scheme
\cite{Neerven3}.

We will focus on the behavior of the cross section at next to leading
order. Although a complete ${\cal O}(\alpha_{\rm s}^2)$ analysis
exists for the total cross section and the rapidity integrated mass
spectrum, the more experimentally useful double-differential cross
section is known only to ${\cal O}(\alpha_{\rm s})$.  The
contributions from soft and virtual gluons, dominant at fixed target
energies and $\tau>~0.01$ \cite{Neerven,Neerven3}, account for only part
of the ${\cal O}(\alpha_{\rm s}^2)$ corrections to ${\rm d}\sigma/{\rm
d}y{\rm d}M^2$ at the higher collider energies.  On the other hand, we
find below that the ${\cal O}(\alpha_{\rm s}^2)$ corrections to
the rapidity integrated cross section are typically quite small for
the kinematic range of interest.  This result supports the reliability
of the ${\cal O}(\alpha_{\rm s})$ prediction from (\ref{eq:nlo})
throughout the rapidity range that contributes most of the cross
section.  Such support is particularly useful in the low mass region
($M\sim M_{J/\psi}$), where a fast convergence of the perturbative
series is far from self evident.

\yoursection{Transverse momentum distributions}

\newcommand{\lb}[1]{\label{#1}\;\;\;\;U\mbox{{\tiny{#1}}}e}
\def\half{{1\over2}}
\def\half{{1\over2}}
\def\etal{{\it et al.}}
\def\d{{\rm d}}
\def\At{A_{\rm\scriptscriptstyle T}}
\def\aw{\alpha_{\rm w}}
\def\tw{\theta_{\rm w}}
\def\as{\alpha_{\rm s}}
\def\hf{{\textstyle{1\over2}}}
\def\dspt{{\d\sigma\over\d\pt^2}}
\def\dspty{{\d\sigma\over\d\pt^2\,\d y\d M^2}}
\def\Dspt{\d\sigma/\d\pt^2}
\def\Dspty{\d\sigma/\d\pt^2\,\d y\d M^2}
\def\ptQ{\left(\pt^2\over Q^2\right)}
\def\Qpt{\left(Q^2\over\pt^2\right)}
\def\Mpt{\left(M^2\over\pt^2\right)}
\def\lnQp{\ln\left(Q^2\over p^2\right)}
\def\lnMp{\ln\left(M^2\over p^2\right)}
\def\asbar{\left(\as\over2\pi\right)}
\def\conv{\circ}
\def\ptcut{\pt^{\rm cut}}
\def\resum{{\rm resum}}
\def\pert{{\rm pert}}
\def\asym{{\rm asym}}
\def\total{{\rm total}}
\def\eulerG{\gamma_{\scriptscriptstyle \rm E}}
\def\Snp{S_{\rm\scriptscriptstyle np}}
\def\Aone{A^{(1)}}
\def\Bone{B^{(1)}}
\def\Atwo{A^{(2)}}
\def\Btwo{B^{(2)}}
\def\Czero{C^{(0)}}
\def\Cone{C^{(1)}}
\def\Abone{{\bar A}^{(1)}}
\def\Bbone{{\bar B}^{(1)}}
\def\Abtwo{{\bar A}^{(2)}}
\def\Bbtwo{{\bar B}^{(2)}}
\def\Czero{C^{(0)}}
\def\Cone{C^{(1)}}
\def\Cbzero{{\bar C}^{(0)}}
\def\Cbone{{\bar C}^{(1)}}
\def\Pone{P^{(1)}}
\def\Ptwo{P^{(2)}}
\def\la{\leftarrow}
\def\bs{b_*}
\def\bmax{b_{\rm\scriptscriptstyle max}}
\def\mw{M_{\rm w}}
\def\mz{M_{\rm z}}
\def\nc{N_{\rm c}}
\def\cf{C_{\rm F}}
\def\tr{T_{\rm R}}
\def\MSbar{$\overline{\rm MS}$}
\def\LambdaQCD{\Lambda_{\rm QCD}}
\def\C{{\vphantom C}^{\phantom (}}

Experiments show that the net transverse momenta of lepton pairs
produced by the Drell-Yan process are of the order of $1$~GeV for a
dimuon mass, $M$, of 10 GeV.  Such values are substantially smaller
than the transverse momenta $\sim M/2$ carried by each of the leptons
individually.  On the other hand, the $\pt$ of a Drell-Yan pair is
much larger than the few-hundred MeV typical of soft QCD.  If we
neglect the transverse momentum of the incoming partons, then the
lowest order process $q \bar q\rightarrow\gamma^*\rightarrow l^+l^-$
produces a final state with net $\pt = 0$.  While any spread in the
initial momentum will increase the final $\pt$ on average, the
intrinsic width of the parton distribution is rather small,
$\langle\pt^2\rangle_{\rm soft}\sim (0.3~{\rm GeV})^2$.  This scale is
determined by the inverse hadron size, since the target and projectile
partons must be localized inside their parent hadrons.  Therefore,
we can attribute part of the measured $\pt$ to the parton's intrinsic
$\pt$, but not all.

The lepton pair acquires additional transverse momentum from
production mechanisms that occur beyond leading order in perturbation
theory \cite{kk,reno}.  For example, in the Compton and annihilation
processes
\begin{equation}
qg \rightarrow q \gamma^*\,\,\,\,\,\,{\rm and}\,\,\,\,\,\,
{\bar q}q \rightarrow \gamma^* g
\label{eq:gluons}
\end{equation}
$\pt$ of the lepton pair can be balanced by the recoil of the
final state quark or gluon.  One can compute the $\pt$ distribution
perturbatively from these processes and their radiative corrections.
The perturbation expansion is well behaved for $\pt\sim M$.
However, at low $\pt$ the expansion breaks down and a
resummation of the perturbation series is required.

To see why this resummation is necessary, observe that the cross
section in the region $\pt^2 \ll M^2$ is dominated by the
leading-logarithm contributions:
\begin{equation}
\dspt \sim {\as\over\pt^2} \ln\Mpt \left[
   v_1 + v_2\as\ln^2\Mpt + v_3\as^2\ln^4\Mpt + \cdots \right] ,
\label{eq:logs}
\end{equation}
where $\as$ is evaluated at the scale $M^2$.  This series is
effectively an expansion in $\as\ln^2(M^2\!/\pt^2)$, rather than $\as$
alone.  The effective expansion parameter can be large at low $\pt$
even if $\as(M^2)$ is small.
The leading-logarithm series (\ref{eq:logs}) describes the effect of
soft gluon radiation from the initial state $q$ and $\bar q$ prior to
their annihilation.  Specifically, these logarithms are remnants of
the mass and collinear singularities arising from the radiated gluons.
The annihilation process in (\ref{eq:gluons}) contributes the term
$\propto \as\ln(M^2\!/\pt^2)/\pt^2$ and, in general, $q \bar{q}
\rightarrow \gamma^* + n~{\rm gluons}$ produces the term of order
$\as^n$.  Fortunately, the coefficients $v_i$ of Eq. (\ref{eq:logs})
are not independent and it is possible to sum the series exactly so
that it applies even when $\as\ln^2(M^2/\pt^2)$ is
large \cite{ddt,cs,css,alta}. In addition, `subleading' logarithm
contributions, though smaller, can also be important.

The formalism needed to sum the leading and subleading logarithms was
developed by Collins, Soper and Sterman \cite{css}.  For each
species of colliding partons, one finds
\begin{equation}
M^2\dspty(\resum) =
     {\pi {\hat\sigma}_0\tau} e_q^2 \int{\d^2b\over(2\pi)^2}
     e^{i\alas{\bf b\cdot p}{T}}
     W(b) ,
\label{eq:resum}
\end{equation}
\begin{eqnarray}
     W(b) &= &
     \exp\left\{
     - \int\nolimits_{\beta^2/b^2}^{M^2} {\d q^2\over q^2}
     \left[ \ln\left({M^2\over q^2}\right)
      A\big(\as(q^2)\big) + B\big(\as(q^2)\big)
     \right] \right\}
  \nonumber \\
  &  & \times~
     (C \conv f_1)(x_1;\beta^2/b^2)
     ~\times~
     (C \conv f_2)(x_2;\beta^2/b^2) .
\label{eq:resum2}
\end{eqnarray}
where $s$ is the total hadronic center-of-mass energy,
$f_1$ and $f_2$ are the projectile and target parton distributions
of the two colliding particles, and
$x_1$ and $x_2$ defined by (4) are the dominant values of $x$ as $\pt
\rightarrow 0$.  Note that the $f_i$ can be $q$, $\overline q$ or $g$,
depending on the process considered.  The integration variable $b$ is
the impact parameter, the variable conjugate to $\pt$, and
$\beta\equiv 2\;{\rm e}^{-\eulerG}$, where $\eulerG$ is Euler's
constant.  To obtain the total Drell--Yan rate at next-to-leading
order, one must sum (\ref{eq:resum},\ref{eq:resum2}) over $q{\bar q}$,
$g{\bar q}$ and $gq$ initial states for all appropriate quark flavors;
see Appendix A in ref.~\cite{russ} for details.  The function $C$ is a
coefficient function that converts the parton distributions $f$ into
distributions $C \conv f$ specific to the process at hand.  The
functions $A$, $B$, and $C(x)$ have perturbative expansions in $\as$,
with $A$ and $B$ starting at order $\as$.  The expansion for $C$
begins at order 1 for quarks and order $\as$ for gluons.  These
functions can be extracted to a given order from the perturbative
result, and have been determined for the Drell--Yan process at
next-to-leading order by Davies {\it et al.} \cite{dav}.

The resummed result (\ref{eq:resum},\ref{eq:resum2}) applies only when
$\pt^2 \ll M^2$ because it includes only those terms that diverge as
$\pt^{-2}$ as $\pt\rightarrow 0$.  Omitted in (\ref{eq:resum}) are
nonsingular contributions that are $\propto \{\pt^2+M^2\}^{-1}$ at
small $\pt$.  At $\pt\sim M$ the singular and nonsingular
contributions become comparable.  On the other hand, conventional
perturbation theory works well at large $\pt$, describing the complete
$\pt$ dependence to a given order in $\as$.

Bridging the low--$\pt$ and perturbative regimes is accomplished by
adding in the terms that are not resummed, the so-called remainder or
nonsingular terms.  Arnold and Kauffman developed a prescription for
calculating the remainder terms that explicitly matches the high and
low $\pt$ results.  Their prescription proceeds as follows.  One first
expands the resummed result (\ref{eq:resum}) in powers of $\as$.  This
series, $\Dspty(\asym)$, contains the singular $1/\pt^2$ part of
complete perturbation series $\Dspty(\pert)$.  We refer to
$\Dspty(\asym)$ as `asymptotic' because it describes the perturbation
series asymptotically as $\pt\rightarrow 0$.  The asymptotic result
in ref.\ \cite{russ} is expressed as convolutions of parton
distributions with the coefficient functions of (\ref{eq:resum2}) and with
Altarelli--Parisi splitting functions arising from the scale dependence
of the parton distributions. With the singular terms isolated in the
asymptotic result, the remainder is the difference between the
perturbative result and the asymptotic result,
\begin{equation}
R =\dspty(\pert) -
\dspty(\asym).
\end{equation}
The perturbation series for the $\pt$ distribution
--- and therefore $R$ ---
has been computed to 2nd order in ref.~\cite{reno}.  The
total cross section is then written
\begin{equation} \dspty(\total) = \dspty(\resum) + \dspty(\pert)
   - \dspty(\asym) .
\label{eq:match}
\end{equation}
The ``matching'' is now manifest: at low $\pt$ the perturbative and
asymptotic pieces cancel, leaving the resummed; at high $\pt$ the
resummed and asymptotic pieces cancel to 2nd order, leaving the
perturbative contribution.  The relative error is explicitly
of order $\as^2$, see ref.~\cite{russ}.

At very high $\pt$ the matching prescription breaks down and one must
switch back to the perturbative result.  This breakdown occurs because
$\Dspty(\asym)$ is only known to 2nd order, while $\Dspty(\resum)$ in
effect contains all orders in $\as$.  For example, $\Dspty(\resum)$
introduces terms $\propto \as^3({\rm ln}\,\pt)^5/\pt^2$ that will not
be cancelled in the 2nd order expression for $\Dspty(\asym)$.
Although such terms are higher order in $\as$ they become important at
large $\pt$ for kinematic reasons.  The resummed and asympotic cross
sections depend on parton distributions evaluated at a fixed $x$,
independent of $\pt$, whereas the parton distributions probed by the
perturbative result fall with increasing $\pt$. Thus, the higher order
terms come to dominate at large $\pt$ and one must switch back to the
perturbative result.  An appropriate value of $\pt$ at which to do
this is when when $\Dspty$ has fallen off to the extent that $R$ is
comparable to the total.  At that point, the terms being resummed no
longer dominate the cross section and at higher $\pt$ the perturbative
prediction is more reliable than (\ref{eq:match}).  The switch is done
at sufficiently high $\pt$ so that the error incurred is free of large
logarithms.

The form factor $W(b)$ contains $\as$ and parton distributions
evaluated at the scale $1/b$, and its evaluation is problematic for $b
> 1~{\rm GeV}^{-1}$.  Moreover, one wishes to include the effect
of the intrinsic $\pt$ of the partons.  Both of these ends are
met by replacing
\begin{equation}
W(b) \rightarrow W(b_*){\rm e}^{-S_{\rm np}(b)}
\end{equation}
where $b_* = b/\sqrt{1 + (b/b_{\rm max})^2}$ and $b_{\rm max} =
0.5$~GeV$^{-1}$.  The Collins-Soper-Sterman formalism
specifies that $S_{\rm np}$ have a term which depends on $\ln M$
and a term which does not and that the $\ln M$ term does not depend on
the colliding hadrons or on the parton $x$'s.  However, beyond these
constraints $S_{\rm np}$ is arbitrary and must be extracted from
experiment.  Ladinsky and Yuan parametrize
\begin{equation}
S_{\rm np} = g_1 b~[b + g_3~{\rm ln}\; (\tau/\tau_1)] +
g_2 b^2~{\rm ln}\; (M/2 M_1),
\label{eq:params}
\end{equation}
where $\tau = x_1x_2$ \cite{lad}.
To fit the ISR $\pt$ distribution
from R209, they take $g_1 = 0.11$~GeV$^2$, $g_2 = 0.58$~GeV$^2$, $g_3
= - 1.5$~GeV$^2$, $\tau_1 = 0.01$ and $M_1=1.6$~GeV.  Note that these
parameter choices are somewhat different from those in ref.\
\cite{russ,dav}.

Momentum distributions presented in the work are computed using a
code adapted from ref.\ \cite{russ}.  One source of uncertainty in
these predictions is the neglect of higher orders in $\as$.  The
difference between the perturbative and matched results at high $\pt$
is one indication of this uncertainty.  Further ambiguity arises in
our estimate of the intrinsic $\pt$ smearing, which is entirely
phenomenological.
%

\yoursection{Angular distributions}

It is possible to probe the spin structure of the production amplitudes
by measuring the angular distribution of the dileptons.

The general form of the angular distribution is
$$
  {{\rm d}
\sigma \over {{\rm d}M^2 {\rm d}y
{\rm d} p^2_{\scriptscriptstyle T} {\rm d} \Omega}} = {3 \over
  16 \pi}
  {{\rm d} \sigma \over { {\rm d} M^2 {\rm d}y
{\rm d} p^2_{\scriptscriptstyle T}}}
\times [1 + \cos^2 \theta
  + {A_0 \over 2} (1- 3\cos^2\theta)
$$
\begin{equation}
  \label{eq:first}
 + A_1 \sin 2 \theta
  \cos\phi + {A_2 \over 2} \sin^2 \theta \cos 2 \phi ]
\end{equation}
\\
where the angles $\theta$ and $\phi$ are measured in the dilepton
rest frame with respect to an arbitrary axis.  For calculations
with underlying QCD processes, it is convenient to
evaluate the $A_i$ in the
Collins--Soper frame \cite{CS}, where the reference axis is the bisector
of the beam and (anti) target directions.  This choice in some
respect minimizes the effect of intrinsic parton transverse
momenta.

For the experimental analysis, it is standard to
use an alternate parameterization
\begin{equation}
  \label{eq:second}
  {{\rm d} \sigma \over {{\rm d} \Omega}} \sim 1 + \lambda \cos^2 \theta +
  \mu \sin 2\theta \cos\phi + {\nu \over 2}  \sin^2\theta \cos 2\phi\,.
\end{equation}
\\
\noindent
The relationship is simply obtained
$$
  \lambda = {{2 - 3A_0} \over { 2+ A_0}}
$$
$$
  \mu = {{2 A_1} \over { 2+ A_0}}
$$
\begin{equation}
  \label{eq:third}
  \nu = {{2 A_2} \over { 2 + A_0}}\,.
\end{equation}
\\
For calculations in perturbative QCD, one imbeds the partonic expressions
for $A_i \times {\rm d}\sigma/{\rm d}M^2 {\rm d}y {\rm d}{p_T}^2 $
into integrals over parton density functions just as in the previous sections.
The Born term involves only zero transverse momentum, and the virtual
photon production amplitude  vanishes for zero helicity.  Thus all of
the $A_i$'s are zero and the angular distribution is purely
$1 + \cos^2\theta$.
For the parton level $A_i$ the leading order (LO) perturbative corrections
of order $\as$ have been calculated through the
spin amplitudes in the annihilation and Compton amplitudes.
One finds in all cases the relationship
$A_0 = A_2$, or equivalently $\lambda = 1 - 2 \nu$, such that the
$\theta$ and $\phi$ distributions are correlated.  Calculations in NLO
of order $\as^2$ are much more complicated \cite{MIRKES1},
but in general only
 alter the angular coefficients at the 10\% level \cite{MIRKES3}.
However, the  correlation above is then violated.

In this study we have calculated the perturbative cross section
 and
amplitudes $A_0$ and $A_1$ using the LO expressions (remember at this level
$A_0 = A_2$).
$$
    {{\rm d} \sigma \over
 {{\rm d} M^2 {\rm d}y {\rm d}p^2_{\scriptscriptstyle T}}} A_i = {{8
    \alpha^2 \tau^2
  \as (\mu)} \over {27 \pi M^6}} \int^1_0 {{dx_1} \over x_1}
  \int^1_0 {{dx_2} \over x_2} \delta (x_1 x_2 - x_1 z_2 - x_2 z_1 + \tau)
$$
$$
  \times \biggl\{ \Bigl[ \sum_k e^2_k
  \bigl( q_k (x_1, \mu) \ \bar q_k (x_2, \mu)
  + (-1)^i (x_1 \leftrightarrow x_2 ) \bigr) \Bigr] \hat{A}_i^{q \bar q}
$$
\begin{equation}
  \label{eq:four}
  + \Bigl[ \sum_k e^2_k \bigr( g (x_1, \mu) (q_k (x_2, \mu) + \bar q_k
  (x_2, \mu)) + (-1)^i (x_1 \leftrightarrow x_2) \bigl) \Bigr] \hat{A}_i^{g q}
  \biggr\},
\end{equation}
\\
\noindent
where $z_{1,2} \equiv [\tau (1 + (\pT/M)^2]^{1 \over 2} e^{\pm y}$
are generalizations of (4) for $\pT \neq 0$.
To calculate the cross section alone, one replaces the parton-level
$\hat{A}_i$ with the parton-level cross section $\hat{\Sigma}$.  Expressions
for these quantities are:

$$
\hat{\Sigma}^{q \bar q} = {{(M^2 - u)^2 + (M^2 - t)^2} \over {ut}}
$$
$$
\hat{A}_0^{q \bar q} = {{M^2 - u} \over {M^2 - t}} + {{M^2 - t} \over
{M^2 - u}}
$$
$$
\hat{A}_1^{q \bar q} = \Bigl[ {{M^2s} \over { ut}} \Bigr]^{1 \over 2}
\bigl({{M^2 - u} \over {M^2-t}} - {{M^2-t} \over
{M^2 - u}}\bigr)
$$
$$
\hat{\Sigma}^{gq} = {{(M^2 - s)^2 + (M^2 - t)^2} \over {-st}}
$$
$$
\hat{A}_0^{gq} = {{-u \bigl[ (s + M^2)^2 + (M^2 -t)^2 \bigr] } \over
{s (M^2 - u) (M^2 -t)}}
$$
\begin{equation}
  \label{eq:fifth}
\hat{A}_1^{gq} = \Bigl[ {{M^2u} \over { st}} \Bigr]^{1 \over 2}
{{(M^2 -u)^2 - 2 (M^2 - t)^2} \over {(M^2 -u) \ (M^2 -t)}}
\end{equation}
\\
\noindent
and one must make the replacement $t \leftrightarrow u$ for
the gluon-quark terms when interchanging projectile and target.

 Note that the invariants $s$, $t$, $u$, and $M^2$ are calculated with
parton momenta
in the  annihilation and
 Compton diagrams, and the $\hat{A}_i$ are given in the
 Collins--Soper frame.  We have used the LO $\as$ values for each
parton distribution set used in these calculations, as is appropriate
for our LO angular distribution expressions.  The scale
is taken to be $\mu = M$  in all cases.

One can see from the structure of parton-level amplitudes in Eq
(\ref{eq:fifth})
how the angular distribution coefficients change as the perturbative
contributions grow with $\pT$.  For the $q \bar q$ subprocess, one
finds the relation

\begin{equation}
  \label{eq:sixth}
\hat{A}_0^{q \bar q} = {{\pT^2} \over {\pT^2 + M^2}} \hat{\Sigma}^{q \bar q}.
\end{equation}
\\
Since this relation holds for all parton momenta, one predicts that
$A_0$ and hence $\lambda$ will be independent of the parton
distribution functions.  It will also be independent of energy and rapidity,
and exhibit a characteristic function of $w \equiv (\pT/M)^2$.
This property was found some time ago \cite{COLLINS,THEWS}, and the
prediction in the Collins--Soper frame at any fixed $y$ is

\begin{equation}
\lambda^{q \bar q} = {{2 - w} \over {2 + 3 w}}.
\end{equation}
\\
One sees that as $w$ increases with $\pT$, the virtual photon
polarization state increases in the zero helicity mode.  The limiting
value as $\pT \rightarrow \infty$ is $\lambda = -1/3$, corresponding
to a factor of two for the ratio of longitudinal to transverse photon
production. There is no corresponding relation for the $A_1$ amplitude.

A similar analysis for the $gq$ amplitudes does not yield a relation
such as Eq.~(\ref{eq:sixth}). However, one can get an approximate
result which only depends on the steeply-rising behavior of the
parton distribution functions at small $x$. If the integral
over parton momenta is saturated by the values at the smallest possible
$x$--values, for small rapidity values one samples only
at the point $-u = -t = \pT^2 + \pT \sqrt{\pT^2 + M^2}$.
The corresponding amplitude relationship is then
\begin{equation}
\hat{A}_0^{gq} \approx {{5 \pT^2} \over {M^2 + 5 \pT^2}} \hat{\Sigma}^{gq},
\end{equation}
\\
which leads to a new characteristic function
\begin{equation}
\lambda^{gq} = {{2 - 5 w} \over {2 + 15 w}}.
\end{equation}
\\
These relations were first found in the Gottfried--Jackson frame \cite{THEWS}
for $y$-integrated quantities, but apply in the form above in the
Collins--Soper frame at fixed $y$ \cite{NA10b} sufficiently small such that
$\hat{A}_1 \approx 0$.
One can see that the characteristic functions are related by a rescaling
of $w$ by a factor of five between the $q \bar q$ and the $gq$ subprocesses.

Our normalized $A_i's$ and calculated $\lambda$,
 $\mu$, and $\nu$ values
are valid only in a region of transverse momentum $\pT$ large
 enough that the
 perturbative terms may be expected to dominate the amplitudes.
 At lower values of $\pT$, the soft gluon resummation technique must be
used to calculate the $\pT$-dependence of the cross section.
As noted by Chiapetta and Le Bellac \cite{CHIAPETTA}, the
 $A_i$ terms do not enter into the resummation, since only the part
 proportional to $1 + \cos^2\theta$ is able to combine with the
 soft gluon resummation amplitude.  Thus at low $\pT$
one should simply replace the
 perturbative cross section with the
 resummed differential cross section, and use this factor to normalize
 the  $A_i$'s integrated over parton distributions.
It is unclear, however, how to determine in general
 where the perturbative region begins.  At the
 Fermilab and CERN fixed-target and ISR energies which provide
  the data presently available for $\pT$ distributions, it appears
that the perturbative terms will dominate only when $\pT > M$.
 On the other hand, calculations for
W and Z production at SPS and Tevatron energies indicate that the perturbative
contributions are dominant already when $\pT \leq M/2.$
Due to this uncertainty, we present for this study only the perturbative
cross section and the perturbative $A_i$ values, plus the calculated
$\lambda$ and $\mu$ values. In regions of small $\pT$, one should use the
resummed cross sections to renormalize the $A_i$ and recalculate the
$\lambda$ and $\mu$ values, but the crossover point in $\pT$ must be
determined independently for each collider energy and dilepton mass.

\yoursection{Nuclear effects}

We now comment on possible nuclear modification in the Drell--Yan
process.  On naive geometrical grounds, one expects that the cross
sections differential in $M$ and $y$ in central ion-ion collisions
increase with nuclear mass by a factor $\propto A^{4/3}$ relative to
the $N-N$ cross section.  Any modification of the parton distributions
in the target and projectile nuclei will modify this dependence.  In
particular, one expects parton shadowing to be very important in the
small $x$ range probed by midrapidity Drell--Yan production at the RHIC
and, especially, at the LHC.  Shadowing can reduce the $A$ dependence
of the cross section relative to the expected increase by as much as a
factor of $\sim A^{1/3}\sim 6$ in Au--Au collisions.  Such a dramatic
suppression would be larger than the combined uncertainties in our
$N-N$ cross section calculations.  It will nevertheless be crucial to
measure the $N-N$ rates at RHIC and LHC to study shadowing and other
such nuclear effects.

Initial state parton scattering has been measured in Drell--Yan studies
of hadron--nucleus collisions.  This scattering does not appreciably
affect rapidity and mass distributions, but can modify the $\pT$
spectrum.  Specifically, initial state scattering broadens the
transverse momentum distribution in a nuclear target relative to a
hadron target, corresponding to an increase $\propto A^{1/3}$ in
$\langle \pT^2\rangle$.  This broadening is measured experimentally.
Note that it is because of this effect that we have
not compared $\pT$ calculations to nuclear target data.

\yoursection{COMPARISON WITH DATA}

In this section we compare calculations to recent experiments in order to
illustrate the level of agreement of the QCD calculations with data.
We have chosen not to optimize the calculations, {\it e.g.,}  by
choosing the scales via some prescription \cite{fudge}.  Instead we vary
the regularization scheme and scales in order to determine the level
of uncertainty in the prediction.  We exclude data on nuclear targets
from our analysis, because nuclear effects are not addressed in this
work.  Even so, our comparisons with data are not exhaustive and we
apologize to our experimental colleagues for our incompleteness.

\yoursection{Mass distributions}

A comparison of the perturbative calculations to the data from fixed--target
experiments is discussed in detail by Rijken and van Neerven
\cite{Neerven3}.  The overall feature of most of the fixed--target data
for ${\rm d}\sigma/{\rm d}M$ is described by the Born term multiplied
by a $K$ factor in the range $1< K <2$.  The reason for this
`factorization' is understood \cite{cs,css}, and the goal of
perturbative calculations of the mass spectrum is to
calculate the $K$ factor.
One finds that the ${\cal O}(\alpha_{\rm s})$
calculation can account for $50-75\%$ of the experimental $K$ factor.
It is not clear whether $K$ can be calculated entirely using
perturbation theory.  As we discuss below the situation improves for
the data at highest energies now available.

In addition Rijken and van Neerven calculate the NNLO,
${\cal O}(\alpha_{\rm s}^2)$,
contributions from soft and virtual gluons ($S+V$) to the
double-differential cross section ${\rm d}\sigma/{\rm d}M{\rm d}\xF$
and study the validity of this approximation at the
${\cal O}(\alpha_{\rm s})$
where the exact result is known.  They find the approximation valid
that at the fixed--target energies for $\sqrt\tau = M/\sqrt s~>~0.3$.
Assuming this to be the case also for the NNLO contribution,
they conclude that part of the discrepancy between the data and the
${\cal O}(\alpha_{\rm s})$ result can be attributed to the $S+V$ contributions
\cite{Neerven3}.

We have extended the comparison in ref.~\cite{Neerven} to the mass
dependence of the double differential cross section, ${\rm
d}\sigma/{\rm d}M{\rm d}\xF$, measured in the FNAL E772 experiment at
800 GeV ($\sqrt s = 38.8$ GeV) \cite{E772} and in the CERN ISR experiment
R209 \cite{CERN-ISR} at $\sqrt s = 44$ and 62 GeV.
In fig.~1 we show the mass distributions from the E772 experiment
\cite{E772} at four different $\xF$ values for the pair,
$\xF =0.125,\ 0.225,\ 0.325$, and 0.425 together with
results from a calculation in the $\overline {\rm MS}$ sceme using the
\MRSDM\ parton distributions \cite{PDFLIB,MRS}.  We take the
scale $\mu$ equal to the mass of the pair, as discussed later.  At low
$\xF$ the data and the perturbative calculation are in
fairly good agreement.  The calculated cross section is slightly below
the data at the lower end of the measured mass range and slightly
above at the higher end.  With increasing $\xF$ the
difference between the data and the calculated results increases at the
low-mass end of the spectrum.

At this energy the validity of the $S+V$ approximation for the ${\cal
O}(\alpha_{\rm s})$ contribution is $\sim 10$ \% at $M=20$ GeV and decreases
to $\sim$~50\% for $M=3$ GeV, the approximate result being larger than
the exact calculation. If the pattern is the same for the second order
corrections, the complete NNLO calculation would deviate from the NLO
results even less than shown in fig.~1.

In figs.\ 2 and 3 the data on ${\rm d}\sigma/
{\rm d}M{\rm d}\xF$ measured at CERN ISR
\cite{CERN-ISR} at $\sqrt s = 44$ and 62 GeV and at $\xF=0$ are compared
to calculations.  At both energies the Born term alone reproduces the
continuum data between the $J/\psi$ and the $\Upsilon$.  For the large
mass region the corrections improve the comparison.  At $\sqrt s=44$
GeV only results for the \MRSDM\ structure functions and
for the scales set to the mass of the pair, $\muF=\muR=M$, are
shown. The NNLO correction calculated in the soft plus virtual gluon
approximation is seen to be clearly smaller than the NLO
correction. Its precise magnitude cannot, however, be trusted with
decreasing values of $\tau=M^2/s$. At $\sqrt s = 62$ GeV the $S+V$
contribution in the NLO term is twice the complete result at small
masses. This implies that the uncertainty in the NNLO correction in
the mass (or $\tau$) range of interest in our extrapolations
to higher energies is of the order of the correction
itself. Fortunately the correction is small, and in the following we
choose to show results with NLO corrections only. We should like to
emphasize that all the available information on the NNLO
contributions, including the full calculation for the rapidity
integrated and total cross sections, indicate that the corrections add
at most 20 \% to the NLO corrected cross sections.

At $\sqrt s = 62$ GeV we show results for \MRSDM, \DZ,
and the GRV HO parton distribution sets
\cite{PDFLIB,MRS,GRV} with $\muF=\muR=M$ and study the scale
dependence in the case of \MRSDM\ set using the NLO
results.  It is not surprising that the different sets give very
similar results since they have been determined from data which covers
or is close to the kinematic region we consider here.  The differences
are too small to discriminate between any of these sets.  Varying the
scale introduces a larger change in the results at this energy.
Specifically, an increase of the scale reduces the calculated result.
Nevertheless, for $M\leq 10$ GeV the change is inconsequential and we
choose to present our extrapolations using $\muF=\muR=M$.

\yoursection{ Transverse momentum distributions}

Transverse momentum spectra computed at next-to-leading order
following Arnold and Kauffman \cite{russ} are compared to data from
ISR experiment R209 $\sqrt{s}=62$~GeV \cite{R209} in
fig.~4.  The nonperturbative parameters employed here
(\ref{eq:params}) were obtained using a leading-order calculation in
ref.~\cite{lad} by fitting data from this experiment and FNAL
experiment E288 at $\sqrt{s}=27.4$~GeV \cite{E288}.  Our NLO
calculations are performed using the \MRSDM\ parton
distributions at the scale $M$.

We compare calculations to Fermilab experiment E772 at $\sqrt{s}=
38.8$~GeV in fig.~5.  The data in fig.~5 are averaged over the range
$0.1< \ala xF <0.3$ for the three different mass bins shown.
Our calculations at this lower energy are in excellent agreement with
the shape of the momentum spectra.  In particular, the variation of
the $\pT$ distributions with mass agrees with
data.  However, present calculations overpredict the integrated rate
by $\sim 50\%$.  In view of this disagreement, we present RHIC and LHC
predictions for transverse momentum distributions normalized to the total
cross section.

\yoursection{Angular distributions}
\def\stacksymbols #1#2#3#4{\def\theguybelow{#2}
\def\verticalposition{\lower#3pt}
\def\spacingwithinsymbol{\baselineskip0pt\lineskip#4pt}
\mathrel{\mathpalette\intermediary#1}}
\def\intermediary#1#2{\verticalposition\vbox{\spacingwithinsymbol
\everycr={}\tabskip0pt
\halign{$\mathsurround0pt#1\hfil##\hfil$\crcr#2\crcr
\theguybelow\crcr}}}
\def\lapproxeq{\stacksymbols{<}{\sim}{2.5}{.2}}
\def\gapproxeq{\stacksymbols{>}{\sim}{3}{.5}}

The only data presently available on the angular distribution
coefficients are from fixed-target $\pi - N$ experiments at Fermilab
E615 \cite{E615} and CERN NA10 \cite{NA10a,NA10b}. These experiments
cover similar kinematic regions, roughly $\sqrt{s} \approx 20$~GeV, $4
\lapproxeq M \lapproxeq 8$~GeV, and $0 \lapproxeq \pT \lapproxeq
3$~GeV.  The general trend of the data produces values of $\lambda$
which are close to unity and almost independent of $\pT$, $\mu$ close
to zero, and $\nu$ increasing with $\pT$.  The perturbative
predictions are in agreement with the $\mu$ and $\nu$ values, but fall
below the $\lambda$ values at the highest $\pT$.  This behavior can be
brought into agreement with data via the procedure of soft gluon
resummation, which also appears necessary to reproduce the magnitude
of the $\pT$ dependent cross section \cite{CHIAPETTA}.  However,
this procedure then brings the predictions for $\nu$ down close to
zero, in significant disagreement with data.  The overall result is a
violation of the relation $1 - \lambda - 2 \nu = 0$ in either the
perturbative or resummed predictions.  This relation should hold
exactly at LO QCD and has slightly positive contributions from the higher
order corrections \cite{MIRKES3}.  The data show definite negative values,
which are difficult to understand in a QCD calculation.  In fact, this
has led to attempts to fit this data with models incorporating initial
state correlations of color fields which lead to spin correlations
\cite{MIRKES2}.  A general conclusion must be drawn from the $\pi -N$
data that the angular distribution results are not well understood
within perturbative QCD.

\yoursection{NUMERICAL RESULTS FOR RHIC AND LHC ENERGIES}

We now turn to our predictions for RHIC and LHC energies and their
uncertainties.

\yoursection{Mass and rapidity distributions}

Mass distributions for p-p collisions are presented in tables 1--4 and
figs.~6--11.
In fig.~6 we show the scale dependence at $\sqrt s = 200$ (a) and  5500 GeV
(b) for different fixed values of the pair mass as a function of $\mu/M$.
Not surprisingly,
the dependence is stronger for smaller masses. The peak at small scale for
$M=4$ GeV is caused by the increase of $\alpha_{\rm s}(M)$ as $M$
approaches the $\alarm{\Lambda}{QCD}$. Perturbative calculations are
not expected to be valid at such a small scale. For large values of the
scale the dependence of the results is weak, although
${\rm d}\sigma/{\rm d}\mu$
does not vanish, as would be the case if $\sigma$ were
locally independent of $\mu$.  We take the scale to be $M$ for our
RHIC and LHC predictions. These results imply that the uncertainty in
these prediction due to the scale ambiguity is $\sim 25\%$.

The scheme dependence of the double differential cross section is shown
in figs.~7 and 8 for $\sqrt s =200$ GeV and 5.5 TeV.  Observe that
the scheme dependence of the parton distributions alone leads to a 10\%
difference in the Born terms.  When the ${\cal O}(\alpha_{\rm s})$
corrections are added the difference between the schemes decreases.
This difference is smaller than the calculated correction, as expected
since the scheme dependence of the cross section is of higher order.  The
difference of the Born terms is expected to be of the order
$\alpha_{\rm s}$.  This seems to be the case even though the
difference is smaller than the ${\cal O}(\alpha_{\rm s})$
corrections.

As mentioned earlier, the ${\cal O}(\alpha_{\rm s}^2)$ corrections to
${\rm d}\sigma/{\rm d}M{\rm d}y$
have recently been studied \cite{Neerven2,Neerven3} but
are not yet completely known. It has been shown \cite{DY1} that at the
present fixed--target energies the ${\cal O}\alpha_{\rm s}$
corrections are dominated by the soft and virtual gluon
corrections. Here we are interested in collisions at larger values of
$\sqrt s$ and smaller masses, down to 2--3 GeV. It seems that the soft
plus virtual gluon approximation breaks down in this domain. However,
the full ${\cal O}(\alpha_{\rm s}^2)$ result is known for the rapidity
integrated cross section ${\rm d}\sigma/{\rm d}M$
\cite{Neerven1}.
%
%
We show the results at the LHC energy, $\sqrt s=5.5$~TeV, both for the
cross section, fig.~9a, and the theoretical $K$ factor fig.~9b. Above
$M=4$ GeV the second order corrections are a small fraction of the first
order corrections and the perturbation theory seems to converge
rapidly. At smaller values of mass the perturbative results become
less reliable but even at $M=2$ GeV the second order correction is not
more than $\sim 10 \%$ of the Born term. It seems that extending the
perturbative calculations down to this mass region is still meaningful
with an uncertainty of $\simlt 25 \%$.

The parton distribution functions are quite well known for $x\simgt
10^{-2}$ and recent parametrizations given by different groups
\cite{PDFLIB} are essentially equivalent.  We give the results at
$\sqrt s =200$ GeV and 5.5 TeV for three different sets: \MRSDZ, \DM\
\cite{MRS}, and GRV HO \cite{GRV}. These
sets differ from each other for $x < 10^{-2}$ and essentially span
the interval compatible with the present HERA data. The \DZ\
set goes slightly below the data and the \MRSDM\ set
slightly above.

Figure~10 shows the mass spectrum for dileptons at RHIC energy.  The
differences in the results for different parton distributions are
$\simlt 20\%$. For the LHC energy the situation is much worse, as shown in
fig.~11. The parton distributions are now probed down to $x=M/\sqrt s \sim
10^{-4}$ and the uncertainty in the cross section at $M=3$ GeV is almost a
factor of 4 decreasing to a factor less than 2 at 10 GeV.

Rapidity distributions at the RHIC energy are presented for fixed pair mass
in fig.~12 for the \MRSDM\ parton distribution set. The
interesting feature is the increase of the cross section at the smaller mass
values as the rapidity increases from 0 to $\sim 3$.
%
%
As is seen from Eq.~(4), $x_1$ increases and
$x_2$ decreases with increasing $y$.  The growth of the cross section
reflects the faster increase of $x_2\bar q(x_2,\muF)$ with decreasing $x_2$
as compared to the decrease of $x_1q(x_1,\muF)$ with increasing $x_1$.
This depends on the detailed shape of the parton distributions at
low $x$ and, {\it e.g.,} for \DZ\ set the cross section is almost
flat in the central rapidity region.

At $\sqrt s = 5.5$ TeV the increase of cross section with $y$ occurs up to
higher values of mass. For $M=3$--10 GeV the cross section peaks at $y\sim 4$
where its value is typically twice that at $y=0$ for the \MRSDM\
set.

\yoursection{Transverse momentum distributions}

Transverse momentum distributions for p-p collisions at the RHIC and
LHC heavy-ion energies are shown in figs.~13--19 normalized to the
$\pT$-integrated cross section.  To understand
some of the features of these spectra, we focus on the RHIC results,
figs.~13--17.
Figure 13 shows $\rho(\pT)$, the normalized $\pT$
distribution calculated at next-to-leading order for $M=4$~GeV and
$y=0$. The normalization factor is the $\pT$ integrated cross section
${\rm d}\sigma/{\rm d}y{\rm d}M $.
The dashed curve is the perturbative prediction valid at high
$\pT$, while the solid thin curve is the matched total
cross section (\ref{eq:match}).  Figure 14 shows the leading order
result at the same energy.  Observe that the difference between the
matched and perturbative curves at high momentum is larger for the LO
calculation compared to the NLO one.

Our prediction --- the thick solid curve in fig.~13 --- switches
between the matched and perturbative solutions, as discussed earlier.
Although the matched result (\ref{eq:match}) formally applies at all
momenta, it is not trustworthy at high $\pT$ where the
remainder $R$ (dash-dotted curve) exceeds the total matched cross
section.  The difference between the matched and perturbative results
is higher order in $\alpha_{\rm s}$; one can regard this difference as
a measure of the uncertainty introduced by our truncation of the
perturbation series.  Observe that this uncertainty is quite small, as
we emphasize in fig.~15 by plotting the results with linear axes.

To illustrate how the matching works, we show the resummed, asymptotic
and perturbative components of the matched solution (13) individually
in fig.~16.  We see explicitly that the divergent asymptotic part
(dash-dotted curve) dominates the perturbation series (thin solid
curve) at low $\pT$.  These contributions cancel
at low $\pT$, so that the matched cross section
is determined by the resummed result (10,11).

In fig.~17 we show the $\pT$ spectrum at RHIC for
$M=10$~GeV.  The effect of switching is smaller at the higher mass
scale.  Figures 18 and 19 show the $\pT$ spectrum
at LHC for $\sqrt{s}=5.5$~TeV, $y=0$ and $M=4$ and 10~GeV at
next-to-leading order.  The matched expression is valid for the entire
region $\pT \le 2M$; switching is unnecessary
in this range.

\yoursection{Angular distributions}

\def\stacksymbols #1#2#3#4{\def\theguybelow{#2}
 \def\verticalposition{\lower#3pt}
 \def\spacingwithinsymbol{\baselineskip0pt\lineskip#4pt}
 \mathrel{\mathpalette\intermediary#1}}
\def\intermediary#1#2{\verticalposition\vbox{\spacingwithinsymbol
 \everycr={}\tabskip0pt
 \halign{$\mathsurround0pt#1\hfil##\hfil$\crcr#2\crcr
  \theguybelow\crcr}}}
\def\lapproxeq{\stacksymbols{<}{\sim}{2.5}{.2}}
\def\gapproxeq{\stacksymbols{>}{\sim}{3}{.5}}

For the calculations of angular coefficients in Eqs.~(16) and (17)
the default parton
distribution functions are the \MRSDM.
We have used fixed-$y$ values mainly
at zero, but also up to maximum allowed by kinematics in some cases.
We study the mass range $3 \leq M \leq 30$ GeV with $0 \leq \pT \leq
2 M$ in each case.\\

Figure~20 shows the $\lambda$ coefficient at RHIC energy for the
default values.  As expected, it decreases with increasing $\pT$ and
approaches a minimum value of $-1/3$ for large $\pT$, and
scales with $\pT/M$ as predicted by
either the $q \bar q$ (exact) or $gq$ (approximate) subprocesses.  The
small scaling violations are an indication that the dominant subprocess
must be $gq$, as one might expect in a p-p interaction.  This is verified
by separate calculation of the subprocess contributions.
We have also verified that the predicted $\lambda$ values are
approximately independent of both $\sqrt{s}$ and the choice of structure
function.\\

All of these calculations were done at $y=0$, where $\mu$ is
consistent with zero, as expected from the target-projectile interchange
symmetry.  At large $y$, however, we expect to see significant
deviations from the simple scaling predictions.  Figure~21
shows the $\lambda$ and $\mu$ values for several rapidities. We see that
as the $\mu$ parameter becomes nonzero, a corresponding nonuniversal
behavior sets in for the $\lambda$ curves.  The corresponding calculations
at LHC energy are shown in fig.~22, where much larger rapidities can be
reached. In fig.~23 we show the corresponding $M$--dependence
at $y = 5$ for LHC.  Clearly, no universal scaling
appears, as exhibited by the same calculations as a function of $\pT/M$ in
fig.~24.

At low $\pT$, all of these calculations will be modified by the
soft gluon resummation procedure.  In general, one would expect
$\lambda\approx 1$ and $\mu \approx 0$
for $\pT$-values up to the point where the perturbative
cross section becomes dominant.  As an example, we calculate $\lambda$ and
$\mu$ at $\sqrt{s} = 38.8$ GeV, where the E772 experiment has measured
the $\pT$ distributions \cite{E772}.  In fig.~25 we compare their data
with the LO perturbative calculations.  As expected, the low-$\pT$ region
shows the perturbative divergences, the intermediate-$\pT$ region is
underestimated by the perturbative terms, and there is some evidence that
the data is being matched by the perturbative calculation as $\pT$ approaches
values near M.  We assume that a proper resummation procedure would
match the data at low-$\pT$ and simply rescale the perturbatively-calculated
$A_i$ with the ratio of measured to perturbative cross sections at each
$\pT$.  Shown in fig.~26 are the $\lambda$ and $\mu$ coefficients for
each case.  One sees that at low-$\pT$ the resummation-corrected values
remain closer to the uncorrected Born term predictions, {\it i.e.,} $\lambda =
1,~\mu = 0.$  Since the $\pT$ values at which the preturbative
calculations become dominant must be separately
determined for each energy and mass value, we simply tabulate the
perturbative cross section and the corresponding $A_0$ and $A_1$ values
for this study at the appropriate RHIC and LHC energies.
For each individual case at low $\pT$, one must then
rescale the $A_i$ with the ratio of perturbative cross section
to resummed (or experimental) cross section values, and then recalculate
the $\lambda$ and $\mu$ parameters.

\yoursection{COMMENTS AND CONCLUSIONS}

We have presented perturbative QCD calculations of the Drell-Yan
process relevent to experiments with heavy ions at future high-energy
colliders.  The applicability of our perturbative calculations
has also been addressed. In the energy
range where experimental results are presently available, the calculations
and the data agree to a level of $\sim 30\%$ or better. In the high energy
domain, $\sqrt s \simgt 200$ GeV, the perturbative series seems to converge
well even down to pair mass of $\sim$2--3 GeV with a NNLO contribution of the
order of 10\% in the rapidity integrated cross section, $\d\sigma/\d M$. The
dependence on the
factorization scheme and on the factorization and renormalization scales is
not
strong except for the smallest considered values of the pair mass, where we
estimate the uncertainty to be $\sim$20--30\%.

At LHC energy the most serious uncertainty arises from the uncertainty in
the parton distribution functions in the small$-x$ region.  Different sets
which are not ruled out by the present HERA data lead to estimates which
differ by a factor of 3--4 for $M\sim 3$ GeV. Since a large pair rapidity
indicates a small $x$ for one of the incoming partons, the uncertainty in
the parton distributions shows up also in the rapidity dependence of pairs.
For the \MRSDZ~set the rapidity distribution is flat in the central region but
for the \MRSDM~it first increases with increasing $y$ before the decrease
at the phase space boundary.

{}From the cross sections for a hard process in a nucleon-nucleon interaction
the number of such processes in a nucleus-nucleus collision can be obtained by
multiplication with the overlap function for the colliding nuclei as defined
in \cite{TAB}. This approach presumes that factorization holds also for
nuclear collisions. It also neglects the dependence of the shadowing of
parton distributions on the local amount of overlap in the tranverse plane.
It should be kept in mind that further studies are needed on the shadowing
and on the validity of the factorization assumption, especially
for this relatively low-mass region of pairs in which we are interested.

\yoursection{Acknowledgements}

We are gratefull to W.L.~van Neerven and P.J.~Rijken for the programs of the
$\d\sigma/\d M$ and $\d\sigma/\d M\d y$ cross sections, to P.L.~McGaughey for
providing the E772 data and to R. Vogt for discussions and helpful comments.

\vskip1cm

\vfil\newpage
\yoursection{Figure Captions}
\begin{description}

\item{1.} The calculated \cite{Neerven3} scaling function $M^3{\rm
d}\sigma/{\rm d}M{\rm d}x$ for four values of Feynman $\xF\equiv
x$ compared to $pp\rightarrow \mu^+\mu^-$ data at $\sqrt s=38.8$ GeV from FNAL
E772 \cite{E772}.  Born, ${\cal O}\as$ and ${\cal O}\as^2$
cross sections are indicated by the dash-dotted, solid and dashed curves.
Next-to-leading corrections are obtained in the $S+V$ approximation.
\item{2.}
Same as fig.~1 compared to ISR R209 data \cite{R209} at $\sqrt{s}= 44$~GeV. %
\item{3.} Same as fig.~1 compared to ISR R209 data \cite{R209} at $\sqrt{s}=
62$~GeV.  Additional curves multiplied by 10 and 100 indicate the dependence
on scale and parton distributions.
\item{4.} The rapidity-integrated
cross section ${\rm d}\sigma/{\rm d}\pT^2$ in the mass range
$5<M<8$~GeV at $\sqrt{s} = 62$~GeV compared to data from CERN R209.  Note that
the normalization of the calculation agrees with that of the data.
\item{5.}
The invariant cross section for $pp\rightarrow \mu^+\mu^-$ at $\sqrt{s}=40$
GeV are compared to measurements from FNAL E772.  The circles, triangles,
squares and the nearby curves represent data and calculations integrated over
the three mass bins $5<M<6$~GeV, $8<M<9$~GeV and $11<M<12$~GeV, respectively.
Data and calculations are averaged over the range $0.1< \xF<0.3$.
Calculations are rescaled by an {\it ad hoc} overall factor of 0.63. 
\item{6.} The scale dependence of the
scaled cross section $M^3{\rm d}\sigma/{\rm d}M{\rm d}y$ at RHIC (a) and LHC
(b) energies in the $\overline{\rm MS}$ scheme at ${\cal O}\as$.
The scales are chosen to be equal, $\muR = \muF=\mu$.\hfill\break
\item{7.} The scheme dependence of the ${\cal O}(\as)$ cross section
at $\sqrt s=200$ GeV for the \MRSDM\ parton distribution set.
\item{8.} Same as in fig.~2 but for the LHC energy, $\sqrt s=5.5$ TeV.
\item{9.} Rapidity integrated cross section ${\rm d}\sigma/{\rm d}M$
(a) and the
theoretical $K$-factor (b) (see text for the definition) at $\sqrt s=5.5$ TeV.
In (a) the dotted curve shows the Born term, dashed curve the
$O(\alpha_{\rm s})$,
and the solid curve the $O(\alpha_{\rm s}^2)$ result.  In (b) the dashed curve
shows the $O(\alpha_{\rm s})$ and the solid curve the $O(\alpha_{\rm s}^2)$
$K$-factor.
\item{10.} The cross section ${\rm d}\sigma/{\rm d}M{\rm d}y$
at $y=0$ as a
function of $M$ for different parton distribution functions for the LHC
energy, $\sqrt s=5.5$ TeV. Dotted curve shows the result for \MRSDM\
set, dash-dotted curve for the GRV HO set and dashed curve for
the \MRSDZ\ set.
\item{11.} Same as fig. 11, except for the RHIC energy, $\sqrt s=200$ GeV.
\item{12.} The cross section ${\rm d}\sigma/{\rm d}M{\rm d}y$
at fixed values of $M$ as a
function of $y$ for the RHIC energy, $\sqrt s=200$ GeV.
\item{13.}  Normalized transverse momentum spectrum $\rho(\pT)=({\rm
d}\sigma/{\rm d}y {\rm d}M{\rm d}\pT^2)/({\rm d}\sigma/{\rm d}y{\rm d}M)$
(thick, solid curve) computed at
next-to-leading order are shown for RHIC at $\sqrt s=200$~GeV, $y=0$,
and $M=4$~GeV.
The dashed curve is the perturbative
prediction valid at high $\pT$, while the solid thin curve
is the matched asymptotic expansion that applies only at low
$\pT$.  The matched solution is not trustworthy at
$\pT$'s where the remainder $R$ (dash-dotted curve) exceeds
to the total matched cross section.
\item{14.} Same as fig.~13, but for a leading order calculation.
All contributions to eq.~(13) are shown explicitly for comparison to
the NLO result in fig.~16.
\item{15.} The cross section ${\rm d}\sigma/{\rm d}y{\rm d}M^2{\rm d}\pT^2$
from fig.~13 plotted without the normalization and with linear axes to
exhibit the true magnitude of the discontinuity incurred by switching.
\item{16.} Same as fig.~13, but showing all the components of the
matched solution individually.
\item{17.}  Normalized $\pT$ spectrum at RHIC for $M=10$~GeV.  The effect
of switching is smaller at the higher mass scale.
\item{18.} Normalized $\pT$ spectrum at LHC for $\sqrt{s}=5.5$~TeV, $y=0$
and $M=4$~GeV at next-to-leading order.  Switching is not necessary for
$\pT \le 2M$.
\item{19.} Same as fig.~18 for $M=10$~GeV.
\item{20.} Angular coefficient $\lambda$ scaling with $\pT/M$.
\item{21.}  Angular coefficients $\lambda$ and $\mu$ variation with rapidity
at $\sqrt{s} = 200$ GeV.
\item{22.} Same as Fig.~21 for $\sqrt{s} = 5500$ GeV.
\item{23.}  Angular coefficients $\lambda$ and $\mu$ variation with $M$ at
large rapidity.
\item{24.}  Angular coefficients $\lambda$ and $\mu$ violation of $M$ scaling
at large rapidity.
\item{25.}  LO perturbative $\pT$ dependence compared with E772 results.
\item{26.}  Angular coefficients $\lambda$ and $\mu$ with resummation
corrections at low $\pT$.
\end{description}
%
\begin{table}[t]
\centerline{
\begin{tabular}{|c|c|c|c|c|c|c|} \hline
\multicolumn{7}{|c|}{ Inclusive cross section for Drell-Yan pairs in p-p
collision } \\
\multicolumn{7}{|c|}{\hbox{$\displaystyle{M^3\frac{{\rm d}\sigma}
{{\rm d}y{\rm d}M}}$} [nb\,GeV$^2$]  } \\
\multicolumn{7}{|c|}{$\sqrt s = 200 $ GeV         } \\
\hline \hline
 & & & & & &  \\
 $M$ [GeV]& Born & Born+LO & $K_{\rm th}$ & Born & Born+LO & $K_{\rm th}$ \\
 & & & & & &  \\
 &\MRSDM&\MRSDM&\MRSDM&\MRSDZ&\MRSDZ&\MRSDZ \\
 & & & & & &  \\
\hline
\hline
  3.0& 0.9694E+01& 0.1338E+02& 1.380& 0.9230E+01& 0.1288E+02& 1.395\\ \hline
  4.0& 0.9523E+01& 0.1274E+02& 1.338& 0.9457E+01& 0.1262E+02& 1.334\\ \hline
  5.0& 0.9304E+01& 0.1222E+02& 1.313& 0.9441E+01& 0.1232E+02& 1.305\\ \hline
  6.0& 0.8946E+01& 0.1163E+02& 1.299& 0.9182E+01& 0.1185E+02& 1.290\\ \hline
  7.0& 0.8525E+01& 0.1100E+02& 1.289& 0.8791E+01& 0.1124E+02& 1.278\\ \hline
  8.0& 0.8091E+01& 0.1037E+02& 1.281& 0.8347E+01& 0.1055E+02& 1.264\\ \hline
  9.0& 0.7703E+01& 0.9836E+01& 1.276& 0.8007E+01& 0.1011E+02& 1.262\\ \hline
 10.0& 0.7304E+01& 0.9297E+01& 1.272& 0.7621E+01& 0.9604E+01& 1.260\\ \hline
 11.0& 0.6920E+01& 0.8793E+01& 1.270& 0.7227E+01& 0.9100E+01& 1.259\\ \hline
 12.0& 0.6551E+01& 0.8310E+01& 1.268& 0.6835E+01& 0.8576E+01& 1.254\\ \hline
 13.0& 0.6196E+01& 0.7861E+01& 1.268& 0.6504E+01& 0.8172E+01& 1.256\\ \hline
 14.0& 0.5854E+01& 0.7439E+01& 1.270& 0.6167E+01& 0.7761E+01& 1.258\\ \hline
 15.0& 0.5530E+01& 0.7028E+01& 1.271& 0.5835E+01& 0.7350E+01& 1.259\\ \hline
 16.0& 0.5222E+01& 0.6643E+01& 1.272& 0.5511E+01& 0.6942E+01& 1.259\\ \hline
 17.0& 0.4924E+01& 0.6274E+01& 1.274& 0.5224E+01& 0.6597E+01& 1.262\\ \hline
 18.0& 0.4641E+01& 0.5924E+01& 1.276& 0.4943E+01& 0.6256E+01& 1.265\\ \hline
 19.0& 0.4373E+01& 0.5584E+01& 1.276& 0.4668E+01& 0.5922E+01& 1.268\\ \hline
 20.0& 0.4120E+01& 0.5273E+01& 1.279& 0.4402E+01& 0.5594E+01& 1.270\\ \hline
\hline
\end{tabular}}
\end{table}
\clearpage

\begin{table}[t]
\centerline{
\begin{tabular}{|c|c|c|c|c|c|c|} \hline
\multicolumn{7}{|c|}{ Inclusive cross section for Drell-Yan pairs in p-p
collision } \\
\multicolumn{7}{|c|}{\hbox{$\displaystyle{M^3\frac{{\rm d}\sigma}
{{\rm d}y{\rm d}M}}$} [nb\,GeV$^2$]  } \\
\multicolumn{7}{|c|}{$\sqrt s = 500 $ GeV         } \\
\hline \hline
 & & & & & &  \\
 $M$ [GeV]& Born & Born+LO & $K_{\rm th}$ & Born & Born+LO & $K_{\rm th}$ \\
 & & & & & &  \\
 &\MRSDM&\MRSDM&\MRSDM&\MRSDZ&\MRSDZ&\MRSDZ \\
 & & & & & &  \\
\hline
\hline
  3.0& 0.1411E+02& 0.1906E+02& 1.350& 0.1125E+02& 0.1582E+02& 1.405\\ \hline
  4.0& 0.1408E+02& 0.1844E+02& 1.309& 0.1234E+02& 0.1652E+02& 1.338\\ \hline
  5.0& 0.1380E+02& 0.1773E+02& 1.284& 0.1282E+02& 0.1666E+02& 1.299\\ \hline
  6.0& 0.1350E+02& 0.1722E+02& 1.276& 0.1297E+02& 0.1662E+02& 1.281\\ \hline
  7.0& 0.1308E+02& 0.1656E+02& 1.265& 0.1288E+02& 0.1630E+02& 1.265\\ \hline
  8.0& 0.1264E+02& 0.1586E+02& 1.254& 0.1268E+02& 0.1585E+02& 1.250\\ \hline
  9.0& 0.1221E+02& 0.1527E+02& 1.250& 0.1242E+02& 0.1545E+02& 1.244\\ \hline
 10.0& 0.1175E+02& 0.1462E+02& 1.243& 0.1209E+02& 0.1494E+02& 1.235\\ \hline
 11.0& 0.1139E+02& 0.1415E+02& 1.242& 0.1180E+02& 0.1455E+02& 1.233\\ \hline
 12.0& 0.1105E+02& 0.1368E+02& 1.238& 0.1151E+02& 0.1414E+02& 1.228\\ \hline
 13.0& 0.1070E+02& 0.1323E+02& 1.236& 0.1119E+02& 0.1373E+02& 1.226\\ \hline
 14.0& 0.1036E+02& 0.1280E+02& 1.235& 0.1086E+02& 0.1331E+02& 1.225\\ \hline
 15.0& 0.1004E+02& 0.1238E+02& 1.233& 0.1054E+02& 0.1290E+02& 1.223\\ \hline
 16.0& 0.9729E+01& 0.1199E+02& 1.232& 0.1022E+02& 0.1250E+02& 1.222\\ \hline
 17.0& 0.9433E+01& 0.1161E+02& 1.231& 0.9914E+01& 0.1211E+02& 1.221\\ \hline
 18.0& 0.9149E+01& 0.1125E+02& 1.229& 0.9611E+01& 0.1172E+02& 1.218\\ \hline
 19.0& 0.8869E+01& 0.1090E+02& 1.228& 0.9306E+01& 0.1133E+02& 1.218\\ \hline
 20.0& 0.8602E+01& 0.1056E+02& 1.227& 0.9011E+01& 0.1095E+02& 1.215\\ \hline
 25.0& 0.7461E+01& 0.9168E+01& 1.228& 0.7844E+01& 0.9564E+01& 1.219\\ \hline
 30.0& 0.6505E+01& 0.8006E+01& 1.230& 0.6797E+01& 0.8292E+01& 1.220\\ \hline
 35.0& 0.5701E+01& 0.7037E+01& 1.234& 0.5976E+01& 0.7328E+01& 1.226\\ \hline
 40.0& 0.5039E+01& 0.6249E+01& 1.240& 0.5262E+01& 0.6481E+01& 1.231\\ \hline
 45.0& 0.4492E+01& 0.5593E+01& 1.245& 0.4704E+01& 0.5830E+01& 1.239\\ \hline
 50.0& 0.4079E+01& 0.5099E+01& 1.250& 0.4261E+01& 0.5305E+01& 1.245\\ \hline
 55.0& 0.3807E+01& 0.4787E+01& 1.257& 0.3972E+01& 0.4978E+01& 1.253\\ \hline
 60.0& 0.3704E+01& 0.4685E+01& 1.264& 0.3862E+01& 0.4872E+01& 1.261\\ \hline
 65.0& 0.3868E+01& 0.4918E+01& 1.271& 0.4027E+01& 0.5114E+01& 1.269\\ \hline
 70.0& 0.4518E+01& 0.5779E+01& 1.279& 0.4693E+01& 0.5998E+01& 1.278\\ \hline
 75.0& 0.6259E+01& 0.8050E+01& 1.286& 0.6487E+01& 0.8349E+01& 1.286\\ \hline
\hline
\end{tabular}}
\end{table}
\clearpage

\begin{table}[t]
\centerline{
\begin{tabular}{|c|c|c|c|c|c|c|} \hline
\multicolumn{7}{|c|}{ Inclusive cross section for Drell-Yan pairs in p-p
collision } \\
\multicolumn{7}{|c|}{\hbox{$\displaystyle{M^3\frac{{\rm d}\sigma}
{{\rm d}y{\rm d}M}}$} [nb\,GeV$^2$]  } \\
\multicolumn{7}{|c|}{$\sqrt s = 5500 $ GeV         } \\
\hline \hline
 & & & & & &  \\
 $M$ [GeV]& Born & Born+LO & $K_{\rm th}$ & Born & Born+LO & $K_{\rm th}$ \\
 & & & & & &  \\
 &\MRSDM&\MRSDM&\MRSDM&\MRSDZ&\MRSDZ&\MRSDZ \\
 & & & & & &  \\
\hline
\hline
  3.0& 0.7467E+02& 0.9717E+02& 1.301& 0.1681E+02& 0.2441E+02& 1.452\\ \hline
  4.0& 0.7381E+02& 0.9077E+02& 1.229& 0.2192E+02& 0.2989E+02& 1.363\\ \hline
  5.0& 0.7200E+02& 0.8650E+02& 1.201& 0.2586E+02& 0.3361E+02& 1.299\\ \hline
  6.0& 0.6993E+02& 0.8427E+02& 1.205& 0.2885E+02& 0.3693E+02& 1.279\\ \hline
  7.0& 0.6757E+02& 0.7984E+02& 1.181& 0.3102E+02& 0.3868E+02& 1.247\\ \hline
  8.0& 0.6522E+02& 0.7636E+02& 1.170& 0.3269E+02& 0.4002E+02& 1.224\\ \hline
  9.0& 0.6305E+02& 0.7415E+02& 1.176& 0.3402E+02& 0.4150E+02& 1.219\\ \hline
 10.0& 0.6074E+02& 0.7109E+02& 1.170& 0.3488E+02& 0.4216E+02& 1.208\\ \hline
 11.0& 0.5866E+02& 0.6842E+02& 1.166& 0.3558E+02& 0.4273E+02& 1.201\\ \hline
 12.0& 0.5695E+02& 0.6677E+02& 1.172& 0.3604E+02& 0.4329E+02& 1.201\\ \hline
 13.0& 0.5529E+02& 0.6442E+02& 1.165& 0.3634E+02& 0.4328E+02& 1.190\\ \hline
 14.0& 0.5364E+02& 0.6262E+02& 1.167& 0.3646E+02& 0.4339E+02& 1.190\\ \hline
 15.0& 0.5211E+02& 0.6055E+02& 1.161& 0.3653E+02& 0.4316E+02& 1.181\\ \hline
 16.0& 0.5070E+02& 0.5904E+02& 1.164& 0.3655E+02& 0.4316E+02& 1.180\\ \hline
 17.0& 0.4939E+02& 0.5733E+02& 1.160& 0.3654E+02& 0.4294E+02& 1.175\\ \hline
 18.0& 0.4816E+02& 0.5575E+02& 1.157& 0.3648E+02& 0.4269E+02& 1.170\\ \hline
 19.0& 0.4691E+02& 0.5421E+02& 1.155& 0.3631E+02& 0.4236E+02& 1.166\\ \hline
 20.0& 0.4575E+02& 0.5285E+02& 1.155& 0.3613E+02& 0.4209E+02& 1.165\\ \hline
\hline
\end{tabular}}
\end{table}
\clearpage

\begin{table}[t]
\centerline{
\begin{tabular}{|c|c|c|c|c|c|c|} \hline
\multicolumn{7}{|c|}{ Inclusive cross section for Drell-Yan pairs in p-p
collision } \\
\multicolumn{7}{|c|}{\hbox{$\displaystyle{M^3\frac{{\rm d}\sigma}
{{\rm d}y{\rm d}M}}$}
[nb\,GeV$^2$]  } \\
\multicolumn{7}{|c|}{$\sqrt s = 14000 $ GeV         } \\
\hline \hline
 & & & & & &  \\
 $M$ [GeV]& Born & Born+LO & $K_{\rm th}$ & Born & Born+LO & $K_{\rm th}$ \\
 & & & & & &  \\
 &\MRSDM&\MRSDM&\MRSDM&\MRSDZ&\MRSDZ&\MRSDZ \\
 & & & & & &  \\
\hline
\hline
  3.0& 0.1671E+03& 0.2167E+03& 1.297& 0.1878E+02& 0.2814E+02& 1.497\\ \hline
  4.0& 0.1659E+03& 0.2073E+03& 1.250& 0.2623E+02& 0.3637E+02& 1.386\\ \hline
  5.0& 0.1616E+03& 0.1918E+03& 1.187& 0.3253E+02& 0.4260E+02& 1.309\\ \hline
  6.0& 0.1564E+03& 0.1797E+03& 1.148& 0.3774E+02& 0.4773E+02& 1.264\\ \hline
  7.0& 0.1508E+03& 0.1642E+03& 1.088& 0.4184E+02& 0.4911E+02& 1.173\\ \hline
  8.0& 0.1454E+03& 0.1632E+03& 1.122& 0.4522E+02& 0.5374E+02& 1.188\\ \hline
  9.0& 0.1405E+03& 0.1612E+03& 1.147& 0.4806E+02& 0.5802E+02& 1.207\\ \hline
 10.0& 0.1355E+03& 0.1534E+03& 1.132& 0.5013E+02& 0.5998E+02& 1.196\\ \hline
 11.0& 0.1309E+03& 0.1459E+03& 1.114& 0.5191E+02& 0.5996E+02& 1.155\\ \hline
 12.0& 0.1269E+03& 0.1468E+03& 1.156& 0.5341E+02& 0.6409E+02& 1.200\\ \hline
 13.0& 0.1231E+03& 0.1406E+03& 1.142& 0.5459E+02& 0.6431E+02& 1.178\\ \hline
 14.0& 0.1193E+03& 0.1374E+03& 1.151& 0.5543E+02& 0.6556E+02& 1.182\\ \hline
 15.0& 0.1160E+03& 0.1333E+03& 1.149& 0.5609E+02& 0.6611E+02& 1.178\\ \hline
 16.0& 0.1129E+03& 0.1295E+03& 1.147& 0.5665E+02& 0.6650E+02& 1.173\\ \hline
 17.0& 0.1100E+03& 0.1249E+03& 1.136& 0.5713E+02& 0.6639E+02& 1.162\\ \hline
 18.0& 0.1072E+03& 0.1209E+03& 1.127& 0.5750E+02& 0.6622E+02& 1.151\\ \hline
 19.0& 0.1045E+03& 0.1186E+03& 1.134& 0.5765E+02& 0.6661E+02& 1.155\\ \hline
 20.0& 0.1019E+03& 0.1146E+03& 1.125& 0.5776E+02& 0.6618E+02& 1.145\\ \hline
 25.0& 0.9120E+02& 0.1037E+03& 1.137& 0.5786E+02& 0.6657E+02& 1.150\\ \hline
 30.0& 0.8283E+02& 0.9371E+02& 1.131& 0.5698E+02& 0.6505E+02& 1.141\\ \hline
 35.0& 0.7664E+02& 0.8638E+02& 1.127& 0.5601E+02& 0.6346E+02& 1.133\\ \hline
 40.0& 0.7179E+02& 0.8131E+02& 1.132& 0.5510E+02& 0.6264E+02& 1.136\\ \hline
 45.0& 0.6848E+02& 0.7750E+02& 1.131& 0.5479E+02& 0.6214E+02& 1.134\\ \hline
 50.0& 0.6693E+02& 0.7538E+02& 1.126& 0.5550E+02& 0.6257E+02& 1.127\\ \hline
 55.0& 0.6752E+02& 0.7622E+02& 1.128& 0.5774E+02& 0.6518E+02& 1.128\\ \hline
 60.0& 0.7175E+02& 0.8103E+02& 1.129& 0.6286E+02& 0.7095E+02& 1.128\\ \hline
 65.0& 0.8248E+02& 0.9345E+02& 1.133& 0.7378E+02& 0.8349E+02& 1.131\\ \hline
 70.0& 0.1068E+03& 0.1207E+03& 1.131& 0.9728E+02& 0.1098E+03& 1.128\\ \hline
 75.0& 0.1648E+03& 0.1864E+03& 1.131& 0.1526E+03& 0.1722E+03& 1.128\\ \hline
 80.0& 0.3334E+03& 0.3776E+03& 1.132& 0.3136E+03& 0.3541E+03& 1.129\\ \hline
\hline
\end{tabular}}
\end{table}
\clearpage

\voffset -1 true in
\begin{table}
\centerline{
\begin{tabular}{|c|c|c|c|c|c|c|} \hline
\multicolumn{7}{|c|}{Angular distribution factors for Drell-Yan pairs in
p-p collision} \\
\multicolumn{7}{|c|}{$\sqrt s = 200 $ GeV, $M = 4 $ GeV} \\
\hline \hline
 & & & & & &  \\
$p_T$ [GeV]& $d\sigma/dM^2d{p_T}^2dy $[GeV$^{-6}$]& $A_0$& $A_1$&
$d\sigma/dM^2d{p_T}^2dy $[GeV$^{-6}$]& $A_0$& $A_1$ \\
&y = 0&  &  &y = 3&  &   \\
& & & & & &  \\
\hline
\hline
0.2& 0.367D-06& 0.0072& 0.0000&	0.170D-06& 0.0062& 0.0140\\ \hline
0.4& 0.772D-07& 0.0296& 0.0000&	0.297D-07& 0.0268& 0.0385\\ \hline
0.6& 0.302D-07& 0.0653& 0.0000&	0.995D-08& 0.0607& 0.0708\\ \hline
0.8& 0.153D-07& 0.1108& 0.0000&	0.435D-08& 0.1038& 0.1082\\ \hline
1.0& 0.888D-08& 0.1623& 0.0000&	0.219D-08& 0.1521& 0.1477\\ \hline
1.2& 0.563D-08& 0.2166& 0.0000&	0.121D-08& 0.2017& 0.1871\\ \hline
1.4& 0.380D-08& 0.2713& 0.0000&	0.710D-09& 0.2502& 0.2245\\ \hline
1.6& 0.267D-08& 0.3247& 0.0000&	0.434D-09& 0.2960& 0.2590\\ \hline
1.8& 0.194D-08& 0.3757& 0.0000&	0.273D-09& 0.3384& 0.2900\\ \hline
2.0& 0.144D-08& 0.4236& 0.0000&	0.175D-09& 0.3770& 0.3173\\ \hline
2.2& 0.110D-08& 0.4683& 0.0000&	0.114D-09& 0.4121& 0.3410\\ \hline
2.4& 0.846D-09& 0.5095& 0.0000&	0.751D-10& 0.4438& 0.3613\\ \hline
2.6& 0.663D-09& 0.5473& 0.0000&	0.498D-10& 0.4726& 0.3786\\ \hline
2.8& 0.525D-09& 0.5820& 0.0000&	0.333D-10& 0.4987& 0.3930\\ \hline
3.0& 0.420D-09& 0.6136& 0.0000&	0.223D-10& 0.5225& 0.4050\\ \hline
3.2& 0.339D-09& 0.6425& 0.0000&	0.149D-10& 0.5444& 0.4148\\ \hline
3.4& 0.276D-09& 0.6688& 0.0000&	0.100D-10& 0.5644& 0.4227\\ \hline
3.6& 0.226D-09& 0.6928& 0.0000&	0.671D-11& 0.5830& 0.4289\\ \hline
3.8& 0.187D-09& 0.7148& 0.0000&	0.448D-11& 0.6002& 0.4337\\ \hline
4.0& 0.155D-09& 0.7347& 0.0000&	0.299D-11& 0.6162& 0.4372\\ \hline
4.2& 0.129D-09& 0.7530& 0.0000&	0.198D-11& 0.6311& 0.4397\\ \hline
4.4& 0.108D-09& 0.7696& 0.0000&	0.131D-11& 0.6452& 0.4412\\ \hline
4.6& 0.915D-10& 0.7849& 0.0000&	0.860D-12& 0.6584& 0.4418\\ \hline
4.8& 0.775D-10& 0.7988& 0.0000&	0.560D-12& 0.6708& 0.4418\\ \hline
5.0& 0.659D-10& 0.8117& 0.0000&	0.362D-12& 0.6826& 0.4411\\ \hline
5.2& 0.563D-10& 0.8235& 0.0000&	0.231D-12& 0.6938& 0.4399\\ \hline
5.4& 0.483D-10& 0.8343& 0.0000&	0.146D-12& 0.7044& 0.4382\\ \hline
5.6& 0.416D-10& 0.8442& 0.0000&	0.910D-13& 0.7145& 0.4361\\ \hline
5.8& 0.359D-10& 0.8534& 0.0000&	0.558D-13& 0.7241& 0.4337\\ \hline
6.0& 0.312D-10& 0.8618& 0.0000&	0.336D-13& 0.7333& 0.4309\\ \hline
6.2& 0.271D-10& 0.8696& 0.0000&	0.198D-13& 0.7421& 0.4279\\ \hline
6.4& 0.236D-10& 0.8768& 0.0000&	0.114D-13& 0.7505& 0.4246\\ \hline
6.6& 0.207D-10& 0.8836& 0.0000&	0.641D-14& 0.7585& 0.4212\\ \hline
6.8& 0.181D-10& 0.8898& 0.0000&	0.348D-14& 0.7662& 0.4177\\ \hline
7.0& 0.159D-10& 0.8955& 0.0000&	0.181D-14& 0.7736& 0.4139\\ \hline
7.2& 0.141D-10& 0.9009& 0.0000&	0.896D-15& 0.7808& 0.4099\\ \hline
7.4& 0.124D-10& 0.9058& 0.0000&	0.416D-15& 0.7879& 0.4058\\ \hline
7.6& 0.110D-10& 0.9105& 0.0000&	0.180D-15& 0.7948& 0.4014\\ \hline
7.8& 0.977D-11& 0.9149& 0.0000&	0.722D-16& 0.8013& 0.3970\\ \hline
8.0& 0.869D-11& 0.9189& 0.0000&	0.268D-16& 0.8073& 0.3929\\ \hline
\hline
\end{tabular}}
\end{table}
\clearpage

\voffset -1 true in
\begin{table}
\centerline{
\begin{tabular}{|c|c|c|c|c|c|c|} \hline
\multicolumn{7}{|c|}{Angular distribution factors for Drell-Yan pairs in
p-p collision} \\
\multicolumn{7}{|c|}{$\sqrt s = 5500 $ GeV, $M = 4 $ GeV} \\
\hline \hline
 & & & & & &  \\
$p_T$ [GeV]& $d\sigma/dM^2d{p_T}^2dy $[GeV$^{-6}$]& $A_0$& $A_1$&
$d\sigma/dM^2d{p_T}^2dy $[GeV$^{-6}$]& $A_0$& $A_1$ \\
&y = 0&  &  &y = 3&  &   \\
& & & & & &  \\
\hline
\hline
0.2& 0.389D-05& 0.0085& 0.0000&	0.545D-05& 0.0081& -0.0071\\ \hline
0.4& 0.844D-06& 0.0341& 0.0000&	0.118D-05& 0.0329& -0.0154\\ \hline
0.6& 0.337D-06& 0.0736& 0.0000&	0.471D-06& 0.0714& -0.0238\\ \hline
0.8& 0.173D-06& 0.1226& 0.0000&	0.241D-06& 0.1194& -0.0317\\ \hline
1.0& 0.101D-06& 0.1768& 0.0000&	0.141D-06& 0.1730& -0.0387\\ \hline
1.2& 0.647D-07& 0.2328& 0.0000&	0.905D-07& 0.2287& -0.0448\\ \hline
1.4& 0.438D-07& 0.2882& 0.0000&	0.614D-07& 0.2841& -0.0498\\ \hline
1.6& 0.309D-07& 0.3415& 0.0000&	0.434D-07& 0.3377& -0.0538\\ \hline
1.8& 0.225D-07& 0.3916& 0.0000&	0.317D-07& 0.3883& -0.0569\\ \hline
2.0& 0.168D-07& 0.4383& 0.0000&	0.237D-07& 0.4357& -0.0592\\ \hline
2.2& 0.128D-07& 0.4813& 0.0000&	0.180D-07& 0.4795& -0.0607\\ \hline
2.4& 0.990D-08& 0.5207& 0.0000&	0.140D-07& 0.5198& -0.0615\\ \hline
2.6& 0.776D-08& 0.5566& 0.0000&	0.110D-07& 0.5567& -0.0618\\ \hline
2.8& 0.615D-08& 0.5895& 0.0000&	0.872D-08& 0.5903& -0.0616\\ \hline
3.0& 0.493D-08& 0.6193& 0.0000&	0.700D-08& 0.6210& -0.0611\\ \hline
3.2& 0.398D-08& 0.6464& 0.0000&	0.567D-08& 0.6489& -0.0602\\ \hline
3.4& 0.325D-08& 0.6712& 0.0000&	0.463D-08& 0.6743& -0.0591\\ \hline
3.6& 0.267D-08& 0.6936& 0.0000&	0.381D-08& 0.6975& -0.0577\\ \hline
3.8& 0.220D-08& 0.7142& 0.0000&	0.315D-08& 0.7186& -0.0562\\ \hline
4.0& 0.183D-08& 0.7329& 0.0000&	0.262D-08& 0.7378& -0.0546\\ \hline
4.2& 0.153D-08& 0.7500& 0.0000&	0.220D-08& 0.7554& -0.0529\\ \hline
4.4& 0.129D-08& 0.7657& 0.0000&	0.185D-08& 0.7714& -0.0512\\ \hline
4.6& 0.109D-08& 0.7801& 0.0000&	0.157D-08& 0.7861& -0.0494\\ \hline
4.8& 0.928D-09& 0.7932& 0.0000&	0.133D-08& 0.7996& -0.0476\\ \hline
5.0& 0.793D-09& 0.8054& 0.0000&	0.114D-08& 0.8120& -0.0457\\ \hline
5.2& 0.680D-09& 0.8166& 0.0000&	0.977D-09& 0.8234& -0.0439\\ \hline
5.4& 0.586D-09& 0.8268& 0.0000&	0.841D-09& 0.8338& -0.0421\\ \hline
5.6& 0.506D-09& 0.8363& 0.0000&	0.727D-09& 0.8434& -0.0403\\ \hline
5.8& 0.439D-09& 0.8450& 0.0000&	0.631D-09& 0.8523& -0.0385\\ \hline
6.0& 0.383D-09& 0.8532& 0.0000&	0.550D-09& 0.8605& -0.0368\\ \hline
6.2& 0.335D-09& 0.8608& 0.0000&	0.480D-09& 0.8681& -0.0351\\ \hline
6.4& 0.293D-09& 0.8677& 0.0000&	0.421D-09& 0.8752& -0.0334\\ \hline
6.6& 0.258D-09& 0.8743& 0.0000&	0.370D-09& 0.8817& -0.0318\\ \hline
6.8& 0.228D-09& 0.8803& 0.0000&	0.326D-09& 0.8878& -0.0302\\ \hline
7.0& 0.202D-09& 0.8860& 0.0000&	0.288D-09& 0.8934& -0.0286\\ \hline
7.2& 0.179D-09& 0.8913& 0.0000&	0.256D-09& 0.8987& -0.0271\\ \hline
7.4& 0.159D-09& 0.8962& 0.0000&	0.227D-09& 0.9036& -0.0256\\ \hline
7.6& 0.142D-09& 0.9008& 0.0000&	0.203D-09& 0.9082& -0.0242\\ \hline
7.8& 0.127D-09& 0.9052& 0.0000&	0.181D-09& 0.9124& -0.0228\\ \hline
8.0& 0.114D-09& 0.9092& 0.0000&	0.162D-09& 0.9165& -0.0214\\ \hline
\hline
\end{tabular}}
\end{table}
\clearpage

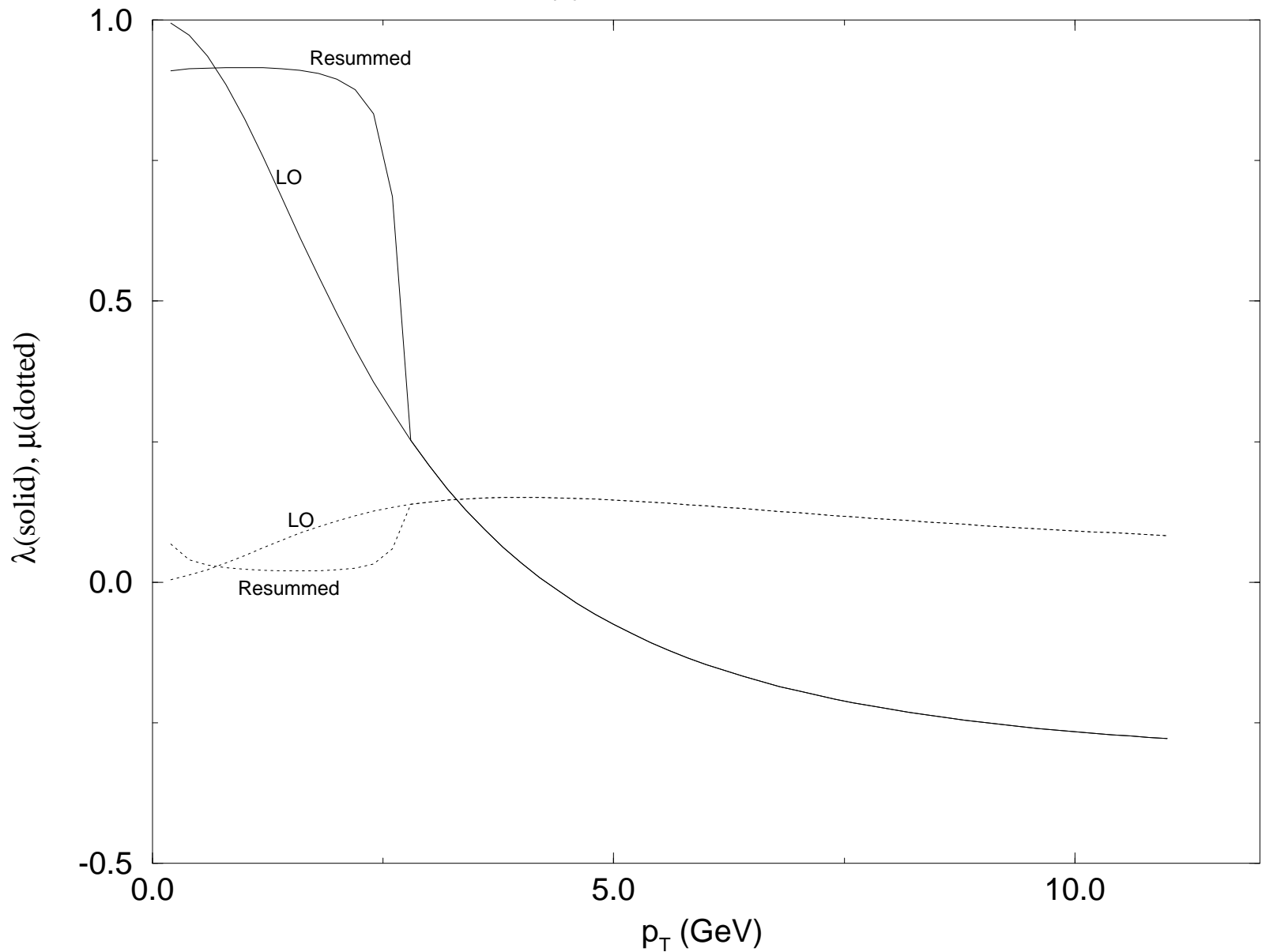
\end{document}